# Empirical assessment of the impact of highway design exceptions on the frequency and severity of vehicle accidents

by

Nataliya V. Malyshkina
School of Civil Engineering
550 Stadium Mall Drive
Purdue University
West Lafayette, IN 47907-2051
nmalyshk@purdue.edu

Fred Mannering
Professor, School of Civil Engineering
550 Stadium Mall Drive
Purdue University
West Lafayette, IN 47907-2051
flm@purdue.edu

**Abstract**

Compliance to standardized highway design criteria is considered essential to ensure the roadway safety. However, for a variety of reasons, situations arise where exceptions to standard-design criteria are requested and accepted after review. This research explores the impact that design exceptions have on the accident severity and accident frequency in Indiana. Data on accidents at roadway sites with and without design exceptions are used to estimate appropriate statistical models for the frequency and severity accidents at these sites using some of the most recent statistical advances with mixing distributions. The results of the modeling process show that presence of approved design exceptions has not had a statistically significant effect on the average frequency or severity of accidents – suggesting that current procedures for granting design exceptions have been sufficiently rigorous to avoid adverse safety impacts.

## Introduction

Design exceptions, which are granted to allow highways to be constructed or reconstructed without meeting all current highway-design standards, have been a focus of concern for many years because the impact of such exceptions, in terms of their effect on road safety, is not well understood. Common reasons for considering design exceptions include: impact to the natural environment; social or right-of-way impacts; preservation of historic or cultural resources; sensitivity to context or accommodating community values; and construction or right-of-way costs (Federal Highway Administration, 1999; American Association of State Highway and Transportation Officials, 2004). Because of the potential of serious safety consequences and tort liability, the process for granting design exceptions is very closely monitored by state and federal highway agencies, although practices and standards for granting design exceptions can vary significantly from state to state (National Cooperative Research Program, 2003).

Over the years, there have been numerous research efforts that have attempted to evaluate the safety impacts of design exceptions. For example, Agent et al. (2002) studied the effect of design exceptions on crash rates in the state of Kentucky. They found that the most common design exception was for a design speed lower than the posted speed limit followed by a lower than standard sight distance, curve radius or shoulder width. With an average of about 39 design exceptions per year in Kentucky, they concluded (based on observations of crash rates) that design exceptions did not result in projects with high crash rates relative to average statewide rates. Unfortunately, in this and many other studies, the amount of data available (which is limited because of the small number of design exceptions granted per year and the highly detailed roadway and accident information required) has made it difficult to develop statistically defensible models to assess the safety impacts of design exceptions in a multivariate framework.

Given the scarcity of design-exception data and associated accident data, some have attempted to infer the effects of design exceptions from statistical models that have been estimated on a simple cross section of roadway segments in an effort to uncover the impact of specific design features (shoulder width, median presence, etc.) on the frequency of accidents and the severity of accidents in terms of resulting injuries. Common statistical approaches to determine the relationship between roadway characteristics and accident frequencies include: Poisson and negative binomial models (Jones et al., 1991; Shankar et al., 1995; Hadi et al., 1995; Poch and Mannering, 1996; Milton and Mannering, 1998; Abdel–Aty and Radwan, 2000; Savolainen and Tarko, 2005; Lord, 2006; Wang and Abdel-Aty, 2008; Lord and Park, 2008); zero–inflated negative binomial models (Shankar et al., 1997; Carson and Mannering, 2001; Lee and Mannering, 2002); negative binomial with random effects models (Shankar et al., 1998); Conway–Maxwell–Poisson generalized linear models (Lord et al., 2008); negative binomial with random parameters (Anastasopoulos and Mannering, 2009) and dual-state negative binomial Markov switching models (Malyshkina et. al, 2009a). For the severity of accidents, quantifying the effects of roadway characteristics on vehicle-occupant injuries have been undertaken using a wide variety of models including multinomial logit models, dual-state multinomial logit models, nested logit models, mixed logit models and ordered probit models (O'Donnell and Connor, 1996; Shankar and Mannering, 1996; Shankar et al., 1996; Duncan et al., 1998; Chang and Mannering, 1999; Carson and Mannering, 2001; Khattak, 2001; Khattak et al., 2002; Kockelman and Kweon, 2002; Lee and Mannering, 2002; Abdel-Aty, 2003; Kweon and Kockelman, 2003; Ulfarsson and Mannering, 2004; Yamamoto and Shankar, 2004; Khorashadi et al., 2005; Lee and Abdel-Aty, 2005; Eluru and Bhat, 2007;Savolainen and Mannering, 2007; Milton et al., 2008; Malyshkina and Mannering, 2009).

However, attempting to infer the impact of design exceptions from general roadway-segment data is potentially problematic because roadway segments that are granted design exceptions are likely to be a non-random sample of the roadway-segment population (segments may have common special features that make them more likely to require a design exception). If this is the case, roadway segments prone to design exceptions will share unobserved effects and the relationship of their characteristics to the frequency and severity of accidents may be significantly different than the relationship on the non-design-exception roadway-segment sample. One way of resolving this problem is to gather a sample of sufficient size that includes roadway segments with design exceptions and similar roadway segments without design exceptions (not a random sample of roadway segments without design exceptions), and to use random parameter models to account for possible unobserved heterogeneity. The intent of this study is to use such a sample and modeling approach to closely assess the effect of design exceptions on the frequency and severity of accidents.

**Empirical Setting**

The Indiana Department of Transportation's highway design criteria are considered essential to ensure the safety of the motoring public. However, for a variety of reasons, situations arise where exceptions to standard-design criteria are requested and accepted after review. Although these decisions are carefully thought out, the safety impacts of various design-criteria exceptions are not well understood. The Indiana Department of Transportation currently has a hierarchy of three levels of highway design criteria. Level One includes those highway design elements which have been judged to be the most critical indicators of highway safety and serviceability. There are 14 Level-One design criteria with minimum standards being met for: design speed; lane widths; shoulder widths; bridge width; bridge structural capacity; horizontal curvature; superelevation transition lengths, stopping-sight distance on horizontal and vertical curves; maximum grade; superelevation rate; minimum vertical clearance; accessibility for the handicapped; and bridge rail safety. Level-Two design criteria are judged to be important to safety and serviceability but are not considered as critical as Level One. Factors in Level Two criteria include: roadside safety elements; the obstruction-free zone; median and side slopes; access control; acceleration lane length; deceleration lane length; shoulder cross slope; auxiliary lane and shoulder widths; minimum grade for drainage; minimum level-of-service criteria; parking lane width; two-way left-turn width; and critical length of grade. Finally, Level Three design criteria include all other design criterion not listed in levels one and two. This research focuses on the impact of design exceptions within the most important Level-One category, which includes the most critical indicators of highway safety and serviceability.

For this study, we consider Level-One design exceptions granted between 1998 and 2003. Our data consist of 35 design exceptions at bridges and 13 on roadway segments. For a control data sample (roadways without design exceptions), 69 control bridges and 26 control roadway segments were carefully chosen so that their characteristics were as similar as possible to those of the design-exception sites (geographic location, road characteristics, traffic conditions, roadway functional classification, and so on). This gives a total of 143 sites.

It is important to note that all bridges are geographically localized sites (they are points on the map). As a result, a procedure was developed for determining the effective length of influence (upstream and downstream of the localized site), and accidents that occurred on this segment of roadway were considered. This segment length was determined to be 1.1 miles (0.55 miles upstream and 0.55 miles downstream of the bridge) using a maximum likelihood estimation as described in Malyshkina et al. (2009b).[1]

Data on individual vehicle accidents were gathered from accident records from 2003 to 2007 inclusive (5 years) from the Indiana Electronic Vehicle Crash Record System. Accident data information included information on weather, pavement conditions, traffic conditions, number and severity of injuries, contributing factors by each vehicle, type and model of each vehicle, posted speed limit, driver's age and gender, safety belt usage, and so on. The data base created from these data included 127 variables for each accident. In all, data on the 5889 accidents that occurred from 2003 to 2007 on the roadway-segment sample were available for our analysis. Of these accidents, 3429 accidents occurred near bridges, 1192 accidents occurred in the proximity of design exception bridges and 2237 accidents occurred in the proximity of control bridges. Among 2460 accidents occurred on roadway segments, 739 accidents occurred on design exception intervals and 1721 accidents occurred on control intervals. Of the 5889 accidents, 26.39% were single-vehicle accidents, 54.54% were two-vehicle accidents involving two passenger vehicles (car, minivan, sport-utility vehicle or pick-up truck), 7.79% were two-vehicle accidents involving a passenger vehicle and a heavy truck, and 11.28% were accidents involving more than two vehicles.

In terms of injury severities, 77.91% of the 5889 accidents were no-injury (property damage only), 21.68% were possible, evident or disabling injury, and 0.41% were fatalities. Among 1931 accidents that occurred on design exception sites, 75.71% were no-injury, 23.82% were injury, and 0.47% were fatalities. Among 3958 accidents occurred on non-design exception (control) sites, 78.97% were no-injury, 20.64% were injury, and 0.39% were fatalities. In terms of accident frequencies, for all 143 roadway segments, the average 5-year accident frequency was 41.13 with a standard deviation of 101.23. Note that for 48 segments with design exceptions, the average 5-year accident frequency was 40.19 with a standard deviation of 90.93. For 95 segments without design

---

[1] To determine the range of influence of bridges on accidents, a multinomial logit model for severity of all accidents is estimated considering all accidents within two miles of the bridge (along the same highway). A distance variable $d_n$ is defined as the distance between the $n^{th}$ accident and the bridge and $D$ is defined as the distance of influence along the roadway segment. A variable is then included in the model where $d_n$ is used only if $d_n < D$, otherwise $D$ is used. Systematically using increasing values of $D$, the severity model that produced the highest log-likelihood was used to determine $D$. The idea is that at some distance away from the bridge the effect of distance from the bridge diminishes. We consider this distance to be extent of influence and create a roadway segment using this value of $D$ so that the total segment length is $2D$ (since the distance is considered in both directions of the roadway).

exceptions, the average 5-year accident frequency was 41.60 with a standard deviation of 106.51.

Detailed information on roadway segments including segment length, locality of the road (rural/urban), number of lanes, median surface type, median width (in feet), interior shoulder presence and width, outside shoulder presence and width, number of bridges, number of horizontal curves, number of ramps, horizontal curve lengths and radii were determined by using the Google Earth software. Average annual daily traffic volumes were obtained from the Indiana Department of Transportation. Road class (interstate, US route, state route, county road, street), rumble strips, median type, road surface type, speed limit value, road type (one-lane, two-lane, multi-lane, one-way, two-way, undivided, divided, alley, private drive) are taken from the available data on individual accident records. Annual accident frequencies on roadway segments were found by matching locations of segments and individual accidents for each of the five years considered (2003-2007 inclusive).

**Methodological Approach – Accident Severity**

For accident injury severity, three possible discrete outcomes are considered: fatal, injury (possible injury, evident injury, disabling injury) and no-injury (property damage only). Past research indicates that the most widely used statistical models to study injury severities have been the multinomial logit model (with nested and mixed logit extensions) and the ordered probit model. However, there are two potential problems with applying ordered probability models to accident severity outcomes (Savolainen and Mannering 2007). The first relates to the fact that non-injury accidents are likely to be under-reported in accident data because they are less likely to be reported to authorities. The presence of under-reporting in an ordered probability model can result in biased and inconsistent model parameter estimates. In contrast, the parameter estimates of the standard multinomial logit model remain consistent in the presence of such under-reporting, except for the constant terms (Washington et. al. 2003). The second problem with ordered probability models is related to undesirable linear restrictions that such models place on influences of the explanatory variables (Eluru et al. 2008; Washington, 2003). As a result of the ordered probit limitations, the multinomial logit approach is used herein.

The standard multinomial logit model with $N$ available data observations and $I$ possible discrete outcomes gives the probability $P_n^{(i)}$ of the $i^{\text{th}}$ outcome in the $n^{\text{th}}$ observation as (see McFadden, 1981)

$$P_n^{(i)} = \frac{exp(\boldsymbol{\beta}_i'\mathbf{X}_{in})}{\sum_{j=1}^{I} exp(\boldsymbol{\beta}_j'\mathbf{X}_{jn})} \quad (1)$$

where $\mathbf{X}_{in}$ is the vector of explanatory variables for the $n^{\text{th}}$ observation and $\boldsymbol{\beta}_i$ is the vector of model parameters to be estimated ($\boldsymbol{\beta}'_i$ is the transpose of $\boldsymbol{\beta}_i$). We use a conventional assumption that the first component of vector $\mathbf{X}_{in}$ is equal to unity, and therefore, the first component of vector $\boldsymbol{\beta}_i$ is the intercept in linear product $\boldsymbol{\beta}'_i\mathbf{X}_{in}$. Note that $P_n^{(i)}$, given by Equation (1), is a valid probability set for $I$ discrete outcomes because the necessary and sufficient conditions $P_n^{(i)} \geq 0$ and $\sum_{i=1}^{I} P_n^{(i)} = 1$ are satisfied[2].

Note that the numerator and denominator of the fraction in Equation (1) can be multiplied by an arbitrary number without any change of the probabilities. As a result, if the vector of explanatory variables does not depend on discrete outcomes (if $\mathbf{X}_{in} = \mathbf{X}_n$), then without any loss of generality one of vectors of model parameters can be set equal to zero. We choose the no-injury vector $\boldsymbol{\beta}_I$ to be zero in this case.

Because accidents are independent events, the likelihood function for the set of probabilities given in Equation (1) is

$$L(\boldsymbol{\beta}_1, \boldsymbol{\beta}_2, \ldots \boldsymbol{\beta}_I) = \prod_{n=1}^{N} \prod_{i=1}^{I} [P_n^{(i)}]^{\delta_{in}} \tag{2}$$

where $\delta_{in}$ is defined to be equal to one if the $i^{\text{th}}$ discrete outcome is observed in the $n^{\text{th}}$ observation and to zero otherwise.

With regard to the magnitude of the influence of specific explanatory variables on the discrete outcome probabilities, elasticities $E_{X_{jn,k}}^{P_n^{(i)}}$ are computed from the partial derivatives of the outcome probabilities for the $n^{\text{th}}$ observation as (see Washington et al. 2003)

$$E_{X_{jn,k}}^{P_n^{(i)}} = \frac{\partial P_n^{(i)}}{\partial X_{jn,k}} \cdot \frac{X_{jn,k}}{P_n^{(i)}} \tag{3}$$

where $P_n^{(i)}$ is the probability of outcome $i$ given by Equation (1), $X_{jn,k}$ is the $k^{\text{th}}$ component of the vector of explanatory variables $\mathbf{X}_{jn}$ that enters the formula for the probability of outcome $j$, and $K$ is the length of this vector. If $j = i$, then the elasticity given by Equation (3) is a direct elasticity, otherwise, if $j \neq i$, then the elasticity is a cross elasticity. The direct elasticity of the outcome probability $P_n^{(i)}$ with respect to variable $X_{in,k}$ measures the percent change in $P_n^{(i)}$ that results from an infinitesimal percentage change in $X_{in,k}$. Note that $X_{in,k}$ enters both the numerator and the denominator of the formula for $P_n^{(i)}$, as given by Equation (1). The cross elasticity of $P_n^{(i)}$ with respect to variable $X_{jn,k}$ measures the percent change in $P_n^{(i)}$ that results from an infinitesimal percentage change in $X_{jn,k}$. Note that $X_{jn,k}$ ($j \neq i$) enters only the denominator of the formula for the probability $P_n^{(i)}$ of the outcome $i$. Thus, cross elasticities measure indirect effects that arise from the fact that the outcome probabilities must sum to unity, $\sum_{j=1}^{I} P_n^{(j)} = 1$. If the absolute value of the computed elasticity $E_{X_{jn,k}}^{P_n^{(i)}}$ of explanatory variable $X_{jn,k}$ is less than unity, then this variable is said to be inelastic, and the resulting percentage change in the outcome probability $P_n^{(i)}$ will be less (in its absolute value) than a percentage change in the variable. Otherwise, the variable is said to be elastic. It is customary to report averaged elasticities, which are the elasticities averaged over all observations. Consider the current study with three discrete outcomes with $i$ = 1, 2, 3 corresponding to the severity levels of fatal, injury and no-injury, the following are the formulas for the averaged direct elasticities (see Washington et al., 2003):

$$\overline{E}_{1;X_k}^{(1)} = \left\langle E_{X_{1n,k}}^{P_n^{(1)}} \right\rangle_n = \left\langle \left[1 - P_n^{(1)}\right] \cdot \beta_{1,k} X_{n,k} \right\rangle_n , \tag{4}$$

$$\overline{E}_{2;X_k}^{(2)} = \left\langle E_{X_{2n,k}}^{P_n^{(2)}} \right\rangle_n = \left\langle \left[1 - P_n^{(2)}\right] \cdot \beta_{2,k} X_{n,k} \right\rangle_n . \tag{5}$$

---

[2] As shown in McFadden (1981), Equation (1) can formally be derived by using a linear function that determines severity probabilities as $S_{in} = \boldsymbol{\beta}'_i\mathbf{X}_{in} + \varepsilon_{in}$, by defining $P_n^{(i)} = Prob\left[S_{in} \geq max_{\forall j \neq i}\left(S_{jn}\right)\right]$ and by choosing the Gumbel (Type I) extreme value distribution for the independently and identically distributed random error terms $\varepsilon_{in}$.

And averaged cross elasticities are

$$\overline{E}_{1;X_k}^{(2)} = \overline{E}_{1;X_k}^{(3)} = \left\langle E_{X_{1n,k}}^{P_n^{(2,3)}} \right\rangle_n = -\left\langle P_n^{(1)} \cdot \beta_{1,k} X_{n,k} \right\rangle_n , \tag{6}$$

$$\overline{E}_{2;X_k}^{(1)} = \overline{E}_{2;X_k}^{(3)} = \left\langle E_{X_{2n,k}}^{P_n^{(1,3)}} \right\rangle_n = -\left\langle P_n^{(2)} \cdot \beta_{2,k} X_{n,k} \right\rangle_n , \tag{7}$$

where brackets $\langle \ldots \rangle_n$ indicate averaging over all observations $n = 1, 2, 3, \ldots, \mathrm{N}$.

The elasticity formulas given above are applicable only when explanatory variable $X_{jn,k}$ used in the outcome probability model is continuous. In the case when $X_{jn,k}$ takes on discrete values, the elasticities given by Equation (3) cannot be calculated, and they are replaced by pseudo-elasticities (for example, see Washington et al., 2003). The later are given by the following equation, which is the discrete counterpart of Equation (3),

$$E_{X_{jn,k}}^{P_n^{(i)}} = \frac{\Delta P_n^{(i)}}{\Delta X_{jn,k}} \cdot \frac{X_{jn,k}}{P_n^{(i)}} \tag{8}$$

where $\Delta P_n^{(i)}$ denotes the resulting discrete change in the probability of outcome $i$ due to discrete change $\Delta X_{jn,k}$ in variable $X_{jn,k}$. For both continuous and discrete variables, because each observation has its own elasticity, only the average across all observations will be reported in the forthcoming empirical analysis.

In addition to simple multinomial logit models, we consider mixed multinomial logit models of accident severity to account for possible variations across observations. In a mixed multinomial logit model, the probability of the $i^{\text{th}}$ outcome in the $n^{\text{th}}$ observation is (see McFadden and Train, 2000; Washington et. al. 2003, page 287)

$$\widetilde{P}_n^{(i)} = \int P_n^{(i)}\, q(\boldsymbol{\beta}_i \,|\, \boldsymbol{\varphi}_i)\, d\boldsymbol{\beta}_i \qquad (9)$$

The right-hand-side of Equation (9) is a mixture of the standard multinomial probabilities $P_n^{(i)}$, given by Equation (1). Probability distribution $q(\boldsymbol{\beta}_i|\boldsymbol{\varphi}_i)$ is the distribution of the multinomial logit parameters $\boldsymbol{\beta}_i$, given fixed parameters $\boldsymbol{\varphi}_i$. The likelihood equation (2) and the elasticity equations (4) through (8) hold for mixed multinomial logit models with $P_n^{(i)}$ replaced by $\widetilde{P}_n^{(i)}$.

## Methodological Approach – Accident Frequency

With regard to accident frequency, the most commonly used statistical models for count data are the Poisson and negative binomial models. The Poisson model is a special case of the more general negative binomial model (a negative binomial model reduces to a Poisson model when the overdispersion parameter is zero). As a result, without loss of generality, we consider only negative binomial models in this study.

The simple standard negative binomial model of accident frequency $A_{tn}$, which is the number of accidents occurred on road segment $n$ during some time period $t$, can be introduced as follows. The probability of $A_{tn}$ is (Washington et al., 2003, page 248)

$$P_{tn}^{(A_{tn})} = \frac{\Gamma(A_{tn} + 1/\alpha)}{\Gamma(1/\alpha) A_{tn}!}\left(\frac{1}{1+\alpha\lambda_{tn}}\right)^{1/\alpha}\left(\frac{\alpha\lambda_{tn}}{1+\alpha\lambda_{tn}}\right)^{A_{tn}}, \qquad \lambda_{tn} = \exp(\boldsymbol{\beta}'\mathbf{X}_{tn}), \qquad (10)$$

where $\lambda_{tn}$ is the mean annual accident rate on roadway segment $n$ during time period $t$, $\mathbf{X}_{tn}$ is the vector of explanatory variables during time period $t$ for the roadway segment $n$, $\Gamma$ is the gamma-function, prime means transpose ($\boldsymbol{\beta}'$ is the transpose of $\boldsymbol{\beta}$). The vector $\boldsymbol{\beta}$ and the over-dispersion parameter $\alpha$ are unknown estimable parameters of the negative binomial model. Because accident events are assumed to be independent, the full likelihood function is,

$$L(\boldsymbol{\beta},\alpha) = \prod_{t=1}^{T}\prod_{n=1}^{N} P_{tn}^{(A_{tn})} \qquad (11)$$

With regard to the magnitude of the influence of specific explanatory variables on the expected accident frequency, instead of the elasticities used for the severity analysis we use marginal effects which are easier to interpret for count-data models. The marginal effect is computed as (see Washington et al., 2003),

$$\frac{\partial E(A_{tn} \mid \mathbf{X}_{tn})}{\partial X_{tn,k}} = \frac{\partial \lambda_{tn}}{\partial X_{tn,k}} = \frac{\partial}{\partial X_{tn,k}} \left[ \exp(\boldsymbol{\beta}' \mathbf{X}_{tn}) \right] = \lambda_{tn} \boldsymbol{\beta} \ , \tag{12}$$

where $X_{tn,k}$ is the $k^{\text{th}}$ component of the vector of explanatory variables $\mathbf{X}_{tn}$. The marginal effect gives the effect that a one unit change in the explanatory variable $X_{tn,k}$ has on the mean accident frequency $\lambda_{tn}$. As was the case with elasticities, because each observation generates its own marginal effect, the average across all observation will be reported in the forthcoming empirical analysis.

The possibility of mixed (random parameters) negative binomial models, which are defined as in the mixed multinomial logit models, is also considered. In a mixed negative binomial model, the probability of $A_{tn}$ accidents occurred on road segment $n$ during annual time period $t$ is (see Greene, 2007; Anastasopoulos and Mannering, 2009)

$$\widetilde{P}_{tn}^{(A_{tn})} = \iint P_{tn}^{(A_{tn})} \ q(\boldsymbol{\beta}, \alpha \mid \boldsymbol{\varphi}) \ d\boldsymbol{\beta} \ d\alpha \tag{13}$$

The right-hand-side of Equation (13) is a mixture of the standard negative binomial probabilities $P_{tn}^{(A_{tn})}$, given by Equation (10). The probability distribution $q(\boldsymbol{\beta},\alpha|\boldsymbol{\varphi})$ is the distribution of the negative binomial parameters $\boldsymbol{\beta}$ and $\alpha$, given fixed parameters $\boldsymbol{\varphi}$. The likelihood equation (equation 11) holds for mixed multinomial negative binomial models with $P_{tn}^{(A_{tn})}$ replaced by $\widetilde{P}_{t,n}^{(A_{tn})}$.

**Methodological Approach – Likelihood Ratio Test**

The likelihood ratio test is generally used in order to determine whether there is a statistically significant difference between models estimated separately for several data bins. In the case when data sample is divided into just two data bins – one which includes data for design exception segments and the other includes data for non-design exception (control) segments, the test statistic is (Washington et al., 2003)

$$X^2 = -2[LL(\boldsymbol{\beta}_{\text{all}}) - LL(\boldsymbol{\beta}_{\text{DE}}) - LL(\boldsymbol{\beta}_{\text{NDE}})], \tag{14}$$

where $LL(\boldsymbol{\beta}_{\text{all}})$ is the model's log-likelihood at convergence of the model estimated on all data, $LL(\boldsymbol{\beta}_{\text{DE}})$ is the log-likelihood at convergence of the model estimated with only sites with design exceptions, and $LL(\boldsymbol{\beta}_{\text{NDE}})$ is the log-likelihood at convergence of the model estimated on sites without design exceptions. If the number of observations is sufficiently large, the test statistic $X^2$, given by Equation (14), is $\chi^2$-distributed with degrees of freedom equal to the summation of parameters estimated in the design exception and non-design exception models minus the number of parameters estimated in the total-data model (Gourieroux and Monfort, 1996).

In the case when the data sample size is small, the asymptotic $\chi^2$ distribution is likely to be a poor approximation for the test statistic $X^2$. To resolve this problem, Monte Carlo simulations can be undertaken to find the true distribution of the test statistic $X^2$. This is done by first generating a large number of artificial data sets under the null hypothesis that the model is the same for segments with and without design exceptions. Then the test statistic values $X^2$, given by Equation (14), for each of the simulated data sets are computed, and these values are used to find the true probability distribution of $X^2$. This distribution is then used for determining the *p*-value that corresponds to the $X^2$ calculated for the actual observed data. The *p*-value is then used for the inference. This Monte-Carlo-simulations-based approach to the likelihood ratio test is universal, it works for any number of observations.

**Estimation results – Accident Severity**

The estimation results for the mixed multinomial logit accident severity model are given in Table 1. The findings in this table show that the severity model has a very good overall fit (McFadden $\rho^2$ statistic above 0.5) and that the parameter estimates are of plausible sign, magnitude and average elasticity. We find that two variables produce random parameters (in the mixed-logit formulation). The indicator variable for having two vehicles involved in the crash was found to be normally distributed in the injury-crash outcome with a mean -1.85 and standard deviation of 2.65. This means that for 75.7% of the observations having two vehicles involved in the crash reduced the probability of the injury outcome and for 24.3% of the observations having two vehicles involved increased the probability of an injury outcome. Also, the parameter for the interstate-highway indicator variable is uniformly distributed with a mean of -2.26 and a standard deviation of 6.03.

Some other interesting results included the age of the at-fault vehicle (where elasticity values show that a 1% increase in at-fault vehicle age increases the probability of a fatal injury by 0.972%) and the age of the oldest vehicle involved in the accident (which also increased the probability of an injury). These two variables may be capturing improvements in safety technologies on newer vehicles.

The presence of snow and slush was found to reduce the probability of fatality and injury, likely due to lower levels of friction which may increase collision time and, therefore, allow energy to be more easily dissipated during a crash. Accidents that did not occur at an intersection and those that occurred in urban areas were less likely to result in an injury (by an average of 12.9% and 21% respectively as indicated by the average elasticities). Finally, accidents involving female drivers who were at fault, having the at-fault vehicle under signal control, having higher posted speed limits, and having driver-related causes indicated as the primary cause of the accident all resulted in a higher likelihood of an injury accident.

Turning to the effect of design exceptions on the severity of accidents, note that in Table 1 the design exception parameter is statistically insignificant, suggesting that design exceptions do not have a statistically significant impact on the severity of accident injuries. To provide further evidence, a likelihood ratio test, described in the previous section, was conducted to determine whether there is a statistically significant difference between mixed multinomial logit models estimated for severity of the accidents that occurred on design exception sites and non-design exception (control) sites. Because the accident-severity data sample is large (5889 accidents), we rely on this $\chi^2$ approximation when doing likelihood ratio test for the accident severity model. The test statistic $X^2$ value, given by Equation (14), was 27.21 with 21 degrees of freedom. The corresponding *p*-value based on the $\chi^2$ distribution, is 0.164 (the critical $\chi^2$ value at the 90% confidence level is 29.62). Therefore, the hypothesis that design exception and non-design exception sites were statistically the same cannot be rejected, and it can be concluded that design exceptions have not had a statistically significant effect on accident severities.[3]

---

[3] Likelihood ratio tests were also conducted to determine if there were differences in accident severities between those roadway segments near bridges and those segments that are not near bridges. To test whether bridge and non-bridge segments were statistically different, we estimated a model on all data and then compared to the separately estimated bridge and non-bridge models. The test statistic is $X^2 = -2[LL(\boldsymbol{\beta}_{\text{all}}) - LL(\boldsymbol{\beta}_{\text{bridge}}) - LL(\boldsymbol{\beta}_{\text{non-bridge}})]$ where $LL(\boldsymbol{\beta}_{\text{all}})$ is the model's log-likelihood at convergence of the model estimated on all data, $LL(\boldsymbol{\beta}_{\text{bridge}})$ is the log-likelihood at convergence of the model estimated with only bridge-segment data, and $LL(\boldsymbol{\beta}_{\text{non-bridge}})$ is the log-likelihood at convergence of the model estimated on non-bridge segment data. This statistic is $\chi^2$ distributed with degrees of freedom equal to the summation of parameters estimated in the bridge and non-bridge models minus the number of parameters estimated in the total-data model. The $X^2$ value of this test was 26.59 with 21 degrees of freedom. The corresponding *p*-value is 0.185 (the critical $\chi^2$ value at the 90% confidence level is 29.62). Therefore, we cannot reject the hypothesis that bridge and non-bridge segments are the same, and, as a result, these segments are considered together.

**Estimation results – Accident Frequency**

With regard to accident frequency, we attempted estimation of a random parameters negative binomial model as shown in Equation (13). Trying various distributions, all estimated parameters were determined to be fixed at the likelihood convergence (standard deviations of parameter estimates across the population were not significantly different from zero implying that the parameters were fixed across observations). Thus, standard negative binomial models are estimated on five-year accident frequencies, and 122 of the 143 road segments had complete data for use in the accident-frequency model estimation. For these 122 road segments, the average 5-year accident frequency was 34.84 with a standard deviation of 65.51.

The negative binomial estimation results are given in Table 2 along with the marginal effects as previously discussed. The results show that the parameter estimates are of plausible sign and magnitude and the overall statistical fit is quite good (McFadden $\rho^2$ statistic above 0.75).

Table 2 shows that the design exception parameter is statistically insignificant again suggesting that design exceptions do not have a statistically significant impact on the frequency of accidents.[4]

Turning to the specific model results shown in Table 2, we find that urban roads have a significantly higher number of accidents and that the higher the degree of curvature (defined as 18000 divided by $\pi$ times the radius of the curve in feet), the lower the accident risk. This second finding seems counterintuitive (sharper curves result in fewer accidents) but this could be reflecting the fact that drivers may be responding to sharp curves by driving more cautiously and/or that such curves are on lower design-speed segments with inherently lower accident risk. Other results in Table 2 show that: increases in average annual daily traffic per lane increase accident frequencies (the marginal effect shows that for every 1000 vehicle increase in AADT per lane the 5-year accident frequency goes up by 2.04 accidents); longer road-segment lengths increase accident frequencies (this is an exposure variable because it is related to the number of miles driven on the roadway segment); and for interstates the higher the number of ramps the greater the number of accidents (with marginal effects indicating that each ramp increases the 5-year accident rate by 6.52 accidents).

The asphalt surface indicator was found to result in fewer accidents. This is likely capturing unobserved information relating to pavement friction and condition (as measured by the International Roughness Index, rutting measurements, and so on) because other studies with detailed pavement-condition information have found the type of roadway surface (concrete or asphalt) to be statistically insignificant (see Anastasopoulos et al., 2008 and Anastasopoulos and Mannering, 2009). Finally, for multilane highways, the presence of an interior shoulder and medians widths of less than 30 feet were found to decrease accident frequency. This latter finding is likely capturing unobserved characteristics associated with highway segments that had medians of 30ft or more (which was about 55% of the sample).

---

[4] The bridge-segment indicator variable was also statistically insignificant suggesting no difference between bridge and non-bridge segments.

We also conducted likelihood ratio tests as was done for the mixed-logit severity analysis. The test statistic $X^2$, given by Equation (14) was 23.00 with 10 degrees of freedom. The corresponding $p$-value based on the $\chi^2$ distribution, is 0.0107 (the critical $\chi^2$ value at the 90% confidence level is 15.99). However, because we have only a limited number of accident-frequency observations (equal to 122), the parameter estimates of the separate frequency models (for design exception and non-design exception segments) are not necessarily statistically reliable (high standard errors) and the asymptotic $\chi^2$ distribution is likely to be a poor approximation for the test statistic $X^2$. To arrive at more defensible results for the likelihood ratio test, Monte Carlo simulations are conducted to determine the true distribution of the test statistic $X^2$, as previously described. These results are shown in Figure 1. In this figure the histogram shows the true distribution of $X^2$ that was obtained from the Monte Carlo simulations ($10^5$ artificial data sets were used) and the solid curve shows the approximate asymptotic $\chi^2$ distribution of $X^2$. The vertical dashed line in this figure shows the $X^2$ value computed for the observed accident-frequency data. The true $p$-value, calculated by using the simulations-based distribution of $X^2$ (this $p$-value is equal to the area of the histogram part located to the right of the dashed line), is 0.0311, which is about three times larger than the approximate $\chi^2$-based value 0.0107. However, both these values are below 5%. Therefore, the hypothesis that design exception and non-design exception sites were statistically the same is rejected, and it can be concluded that design exceptions have a statistically significant effect on accident frequencies. This is an extremely important finding. The fact that the indicator variable for design exceptions was found to be statistically insignificant suggests that the difference between design exception and non-design exception segments in terms of higher accident frequencies is not significant. However, the likelihood ratio test results suggest that the process (estimated parameters) generating accident frequencies of the design exception and non-design exception segments are significantly different. This has important implications in that potential changes in explanatory variables **X** could produce significantly different accidents frequencies between design exception and non-design exception segments. While more data would be needed to completely uncover these effects, this finding indicates that caution needs to be exercised even when granting design exceptions that appear to have been acceptable based on historical data.

**Summary and Conclusions**

Overall, our results suggest that the current process used to grant design exceptions has been sufficiently strict to avoid adverse safety consequences resulting from design exceptions – although the finding that different processes may be generating the frequencies of accidents in design exception and non-design exception segments is cause for concern with regard to future granting of design exceptions.

Our specific findings (even with the limited data available to us) provide some insight into areas where caution should be exercised when granting Level One design exceptions. With regard to the severity of accidents, while most of the factors that affected severity were driver characteristics, we did find that urban-area accidents have a lower likelihood of injury and that the posted speed limit is critical (higher speed limits result in a significantly higher probability of an injury accident). Thus, urban/rural location and design exceptions on highways with higher speed limits need to be given careful scrutiny.

With regard to the frequency of accidents, we find that horizontal curvature is critical and thus special attention needs to be paid to design exceptions relating to horizontal curves. For multilane highways, the presence of interior shoulders was found to significantly reduce the frequency of accidents so this should be considered carefully when granting design exceptions. Also, higher accident frequencies were found in urban areas suggesting that special attention should be given to design exceptions that could compromise safety in these areas (as expected, urban areas have higher accident frequencies but lower severities). Finally, the asphalt-surface indicator was found to result in fewer accidents. As stated previously, this is likely capturing unobserved information relating to pavement friction and condition (as measured by the International Roughness Index, rutting measurements, and so on), and suggests that friction and pavement conditions have to be watched closely when design exceptions are granted.

In terms of a process in the form of a decision support system for guiding future Level One design exceptions, the statistical findings of this research effort suggest that using previous design exceptions as precedents would be a good starting point. While the current study indicates that the design exceptions granted over the 1998-2003 timeframe have not adversely affected overall safety, the number of available design exceptions is too small to make broad statements with regard to policy. Thus, a case by case comparison with previously granted design exceptions is the only course of action that can be recommended.

Table 1. Estimation results for the mixed multinomial logit model of accident severities.

| Variable | Parameter (*t*-ratio) | | Averaged elasticities* | | | | | |
|---|---|---|---|---|---|---|---|---|
| | **fatal** | **injury** | $\overline{E}_1^{(1)}$ | $\overline{E}_1^{(2)}$ | $\overline{E}_1^{(3)}$ | $\overline{E}_2^{(1)}$ | $\overline{E}_2^{(2)}$ | $\overline{E}_2^{(3)}$ |
| **Fixed parameters** | | | | | | | | |
| Constant | -6.09 (-10.4) | -4.59 (-7.26) | - | - | - | - | - | - |
| Two vehicle indicator (1 if two vehicles are involved, 0 otherwise) | -2.41 (-3.63) | - | -1.45 | .0012 | .0030 | - | - | - |
| Snow-slush indicator (1 if roadway surface was covered by snow or slush, 0 otherwise) | -.843 (-2.41) | -.843 (-2.41) | -.0460 | .0001 | .0002 | .0038 | -.0255 | .0038 |
| Help indicator (1 if help arrived in 10 minutes or less after the crash, 0 otherwise) | .609 (3.69) | .609 (3.69) | .358 | -.0008 | -.0014 | -.0488 | .1576 | -.0488 |
| Number of occupants in the vehicle at fault | -.328 (-2.54) | -.328 (-2.54) | -.479 | .0010 | .0016 | .0574 | -.2301 | .0574 |
| The largest number of occupants in any single vehicle involved in the crash | .303 (2.70) | .303 (2.70) | .526 | -.0013 | -.0040 | -.0666 | .243 | -.0666 |
| Age of the vehicle at fault (in years) | .139 (3.38) | - | .972 | -.0033 | -.0055 | - | - | - |
| Age of the oldest vehicle involved in the accident (in years) | - | .0417 (2.96) | - | - | - | -.0457 | .160 | -.0457 |
| Non-intersection indicator(1 if the accident did not occur at an intersection, 0 otherwise) | - | -.409 (-2.43) | - | - | - | .0285 | -.129 | .0285 |
| Indiana vehicle license/fault indicator (1 if the at fault vehicle was licensed in Indiana, 0 otherwise) | - | .390 (2.11) | - | - | - | -.0390 | .145 | -.0390 |
| Urban indicator (1 if the accident occurred in an urban location, 0 otherwise) | - | -.686 (-2.78) | - | - | - | .0533 | -.210 | .0533 |
| Driver/cause indicator (1 if the primary cause of accident is driver-related, 0 otherwise) | - | 2.27 (8.28) | - | - | - | -.255 | .814 | -.255 |
| Posted speed limit (if the same for all vehicles involved) | - | .0239 (2.66) | - | - | - | -.129 | .511 | -.129 |
| Signal/fault indicator (1 if the traffic control device for the vehicle at fault is a signal, 0 otherwise) | - | .724 (2.90) | - | - | - | -.0146 | .0353 | -.0146 |

Table 1. (Continued)

| Variable | Parameter (*t*-ratio) | | Averaged elasticities* | | | | | |
|---|---|---|---|---|---|---|---|---|
| | **fatal** | **injury** | $\overline{E}_1^{(1)}$ | $\overline{E}_1^{(2)}$ | $\overline{E}_1^{(3)}$ | $\overline{E}_2^{(1)}$ | $\overline{E}_2^{(2)}$ | $\overline{E}_2^{(3)}$ |
| Female-fault indicator (1 if gender of the driver at fault ids female, 0 otherwise) | - | .308 (2.06) | - | - | - | -.0155 | .0543 | -.0155 |
| Middle age indicator (1 if the age of the oldest driver is between 30 and 39 years old, 0 otherwise) | - | .577 (3.12) | - | - | - | -.0162 | .0501 | -.0162 |
| **Design exception and segment-type parameters** | | | | | | | | |
| Design exception indicator (1 if the road segment had a design exception, 0 otherwise) | .460 (.752) | -.0974 (-.622) | .147 | -.0004 | -.0007 | .0038 | -.0149 | .0038 |
| Bridge-segment indicator (1 if the road segment is a bridge segment, 0 otherwise) | -.244 (-.395) | .175 (1.11) | -.141 | .0003 | .0005 | -.0119 | .0443 | -.0119 |
| **Random parameters** | | | | | | | | |
| Two vehicle indicator (1 if two vehicles are involved, 0 otherwise) | - | -1.85 (-3.67) | - | - | - | .0223 | -.0185 | .0223 |
| Interstate indicator (1 if the roadway segment is an interstate highway, 0 otherwise) | - | -2.26 (-2.59) | - | - | - | -.0141 | .104 | -.0141 |
| **Standard deviations of parameter distributions** | | | | | | | | |
| Two vehicle indicator (1 if two vehicles are involved, 0 otherwise) | - | 2.65 (3.86) normal | - | - | - | - | - | - |
| Interstate indicator (1 if the roadway segment is an interstate highway, 0 otherwise) | - | 6.03 (3.74) uniform | - | - | - | - | - | - |
| Log-likelihood at convergence | -1828.38 | | | | | | | |
| Restricted log-likelihood | -4027.51 | | | | | | | |
| Number of parameters | 21 | | | | | | | |
| Number of observations | 3666 | | | | | | | |
| McFadden $\rho^2$ | 0.546 | | | | | | | |

[*] Refer to equations (3)–(6), where subscript and superscript outcomes "1", "2", "3" correspond to "fatal", "injury", "no-injury". The subscript represents the outcome that is being considered for X. The superscript represents the outcome that is being considered for the change in probability. If subscript and superscript values are the same, the elasticity is a direct elasticity estimating the effect of a change in variable X in the subscript outcome has on the probability of that same outcome. If subscript and superscript values differ, they are cross elasticities estimating the effect of a change in variable X in the subscript outcome has on the probability of the superscript outcome.

Table 2. Estimation results for the standard negative binomial model of 5-year accident frequencies

| **Variable** | **Parameter (*t*-ratio)** | **Averaged marginal effect** |
|---|---|---|
| Constant | 3.12 (7.23) | - |
| Urban indicator (1 if the accident occurred in an urban location, 0 otherwise) | 1.80 (4.43) | 71.9 |
| Degree of curvature of the sharpest horizontal curve on the road segment | -.0562 (-2.08) | -2.24 |
| Average annual daily traffic per lane in thousands | .0509 (2.28) | 2.04 |
| Natural logarithm roadway segment length in miles | .937 (2.83) | 37.5 |
| Total number of ramps on road segment if interstate | .163 (2.00) | 6.52 |
| Asphalt surface indicator (1 if roadway surface is asphalt, 0 otherwise) | -1.08 (-3.13) | -43.4 |
| Multilane highway interior shoulder indicator (1 if road segment is a divided highway with an interior shoulder, 0 otherwise) | -1.25 (-3.10) | -50.1 |
| Multilane highway median-width indicator (1 if median width is less than 30 feet, 0 otherwise) | -.905 (-2.55) | -36.2 |
| **Design exception and segment-type parameters** | | |
| Design exception indicator (1 if the road segment had a design exception, 0 otherwise) | .0601 (0.204) | - |
| Bridge-segment indicator (1 if the road segment is a bridge segment, 0 otherwise) | -.155 (-0.414) | - |
| **Dispersion parameter** | | |
| Over-dispersion parameter (α) | 1.37 (7.94) | - |
| **Model statistics** | | |
| Log-likelihood at convergence | -1963.29 | |
| Log-likelihood at zero | -472.77 | |
| Number of parameters | 12 | |
| Number of observations | 122 | |
| McFadden $\rho^2$ | 0.759 | |

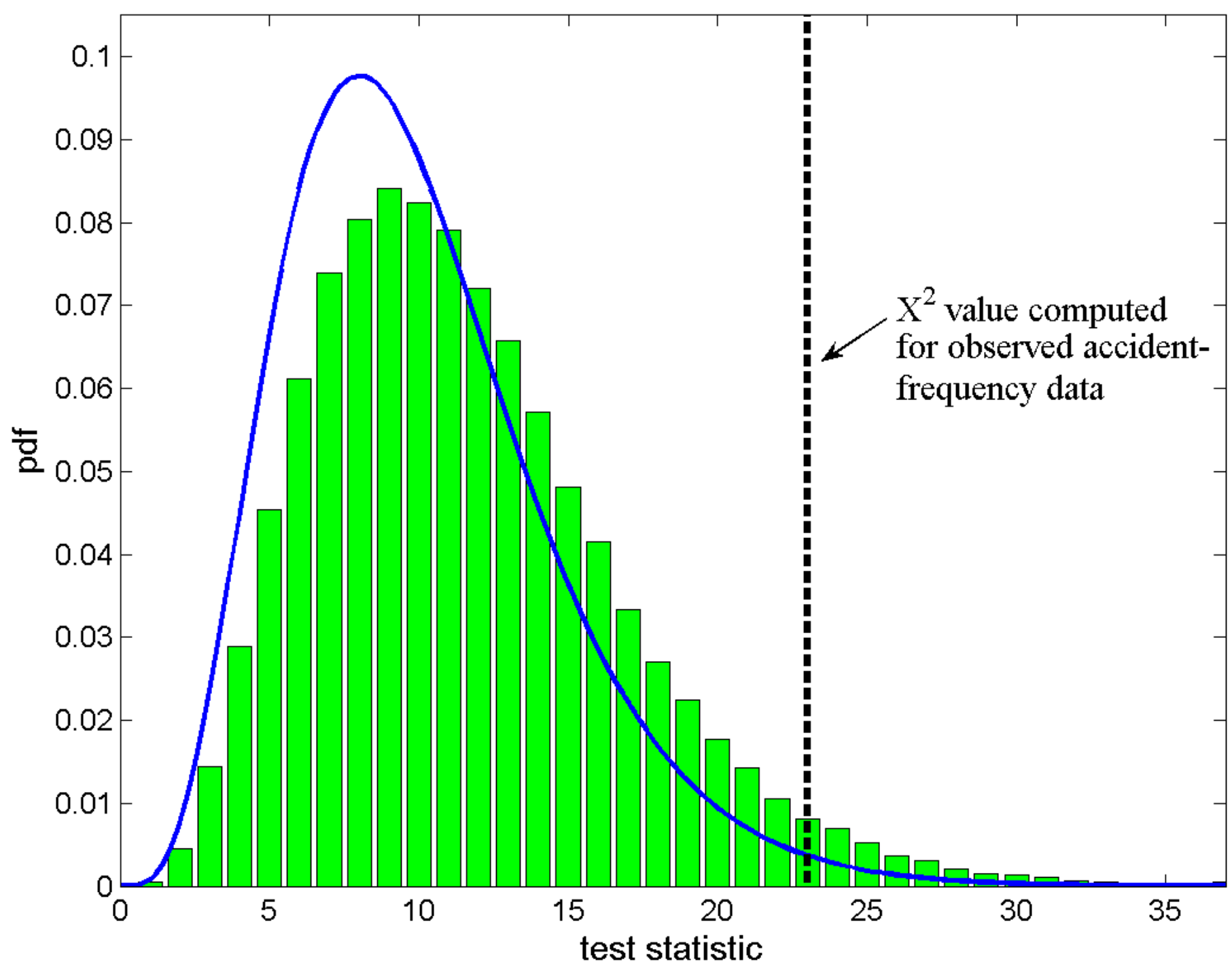

Figure 1. The $X^2$ test statistic distribution for accident frequency (histogram shows the true distribution of $X^2$ from the Monte Carlo simulations and the solid curve shows the approximate asymptotic $\chi^2$ distribution of $X^2$).